\documentclass[english,prd,twocolumn,epsf,superscriptaddress,nofootinbib,preprintnumbers]{revtex4}
\usepackage[latin1]{inputenc}
\usepackage{graphicx, psfrag}
\usepackage{amssymb}
\usepackage[colorlinks=true, citecolor=blue, urlcolor = blue, linkcolor= red, bookmarks=true]{hyperref}
\usepackage{float}
\usepackage{amsmath}
\usepackage{amsfonts}
\usepackage{dcolumn}
\usepackage{hyperref}
\usepackage{subfigure}
\usepackage{pgfplots}
\usepackage{epstopdf}
\begin{document}
\def \beq{\begin{equation}}
\def \eeq{\end{equation}}
\def \bse{\begin{subequations}}
\def \ese{\end{subequations}}
\def \bea{\begin{eqnarray}}
\def \eea{\end{eqnarray}}
\def \bem{\begin{displaymath}}
\def \eem{\end{displaymath}}
\def \bem{\begin{pmatrix}}
\def \eem{\end{pmatrix}}
\def \beb{\begin{bmatrix}}
\def \eeb{\end{bmatrix}}
\def \bc{\begin{center}}
\def \ec{\end{center}}
\def \bb{\bibitem}
\def \bs{\boldsymbol}
\def \nn{\nonumber}

\newcommand{\cc}{\color{red}}
\newcommand{\cb}{\color{blue}}

\title{Accretion onto a noncommutative geometry inspired black hole}

\author{Rahul Kumar}\email{rahul.phy3@gmail.com}
\affiliation{Centre for Theoretical Physics,
 Jamia Millia Islamia,  New Delhi 110025
 India}

\author{Sushant G. Ghosh}\email{sghosh2@jmi.ac.in}
\affiliation{Centre for Theoretical Physics, Jamia Millia Islamia,  New Delhi 110025 India}
\affiliation{Multidisciplinary Centre for Advanced Research and Studies (MCARS),\\ Jamia Millia Islamia, New Delhi 110025, India}
\affiliation{Astrophysics and Cosmology Research Unit,
	School of Mathematics, Statistics and Computer Science,
	University of KwaZulu-Natal, Private Bag X54001,
	Durban 4000, South Africa}

\begin{abstract}
The spherically symmetric accretion onto a noncommutative (NC) inspired Schwarzschild black hole is treated for a polytropic fluid. The critical accretion rate $\dot{M}$, sonic speed $a_s$ and other flow parameters are generalized for the NC inspired static black hole and compared with the results obtained for the standard Schwarzschild black holes. Also explicit expressions for gas compression ratios and temperature profiles below the accretion radius and at the event horizon are derived. This analysis is a generalization of Michel's solution to the NC geometry. Owing to the NC corrected black hole, the accretion flow parameters also have been modified.  It turns out that $ \dot{M} \approx {M^2}$ is still achievable but $r_s$ seems to be substantially decreased due to the NC effects. They in turn do affect the accretion process.    
\end{abstract}

\maketitle

\section{Introduction}
 The process by which compact massive astronomical objects capture the ambient matter present in the interstellar medium that leads to an increase of their mass (and possibly their angular momentum also) is called \textit{accretion} \cite{Ignacio}. It is one of the most ubiquitous processes in
astrophysics. This accretion of gas by compact objects is likely to be the source of the energy of observed X-rays binary, quasar and active galactic nuclei. During the accretion, the kinetic energy of infalling gas increase on the essence of gravitational energy and subsequently results in the increase of radiant energy of the object. The history of the research on the accretion of an ideal fluid onto a compact object begins with the work of Hoyle and Lyttleton~\cite{Hoyle1}, who studied the effects of accretion onto the change in terrestrial climate and, subsequently, the considerable change in the radiation emitted. In their work they neglect pressure effects. Later in this extension Bondi \cite{Bondi} in his pioneering work studied the stationary, spherically symmetric accretion of non-relativistic gas onto compact objects. This work was also the generalization of Bondi'S and Hoyle'S earlier work \cite{Bondi1}. It may be noted that Bondi's work was distinct from Hoyle and Lyttleton's \cite{Hoyle1} work in the sense that in the former case accreting gas was at rest at infinity, while in the later case it has finite velocity. So far the entire work was done using Newtonian gravity. A relativistic generalization was made by Michel \cite{Michel} (see also \cite{keylist} for further generalizations and supplements to Michel's solution) who studied the stationary, spherically symmetric and relativistic accretion onto the compact objects.  Several generalizations to Michel's work have been suggested  to discuss accretion in various other forms \cite{keylist,stbook}, viz., properties of accretion onto the Schwarzschild black hole have been broadly investigated in \cite{perfect, perfect1, perfect2, perfect3} and Begelman \cite{Begelman} examined some aspects of the critical points of the accretion problem. Other studies include the accretion onto a higher dimensional black hole \cite{John, Sharif11}, on a black hole in a string cloud background in \cite{Apratim}, on a charged black hole \cite{Jamil, charge1, charge} and on a rotating black hole \cite{shap73b, shap74} (see an overall review \cite{Carr:2010wk}). A study of general relativistic spherical accretion with and without back-reaction has also been done in \cite{malec}. The result showed that the mass accretion rate is enhanced in the absence of a back-reaction. \newline The aim of this work is to explicitly bring out the NC geometrical effects on the spherically symmetric accretion onto Schwarzschild black holes. It is well known that the accretion process is a powerful indicator of the physical nature of the central celestial objects, which means that the analysis  of the accretion around the nonrotating NC inspired black hole can help us to understand  the regime where general relativity breaks down. 

However, even after intensive research on black holes some aspects still require cogent interpretation, as the present description of the late stage black hole evaporation is still inadequate \cite{Hawking1}. However, we can explain black hole evaporation as a semiclassical process where the breakdown happens in the limit $M_{BH}< M_{pl}$ during evaporation. Furthermore, the black hole temperature ($T_\mathrm{H} \approx 1/M $) diverges in the last stage of the evaporation process, that is $T_H \rightarrow \infty $ as $M\rightarrow 0$. However, it is expected that this divergence will not actually occur, because in the Planck phase the black hole will be dramatically disturbed by strong quantum gravitational effects under $M_{BH}<M_{pl}$. To efficiently describe all evaporation phases of a black hole one must include the quantum field theory (QFT) effects in curved spacetime and ensure the controlled behavior of the theory at very high energies \cite{Padmanabhan1}. Noncommutativity (NCY) is one of the promising ways to approach QFT in curved spacetime \citep{Smailagic1, Smailagic2, Smailagic3, Chaichian1}, which arises naturally also in string theory \cite{A. Connes, 3}. It has been widely accepted that quantum gravity must entail an uncertainty principle which prevents the exact position measurement below the Planck scale \cite{Snyder}.  The NC geometry is naturally encoded in the commutator 
\beq
[x^{\mu},x^{\nu}]=i\theta^{\mu\nu},\quad [\partial_{\mu},\partial_{\nu}]=0,
\eeq
where $\theta^{\mu\nu}$ is an anti-symmetric real matrix which determines the discretization of the spacetime. This commutator leads toward the resulting uncertainty relation in the quantum gravity regime,
\beq
\Delta x^{\mu}\Delta x^{\nu}\geq \frac{1}{2}|\theta^{\mu\nu}|.
\eeq
Further, the Lorentz invariance, unitarity and UV finiteness of NCQFT is ensured by assuming an unique parameter $\theta$ expressing the spacetime NCY. Consequently, we must have the same NC parameter in all the different planes. Therefore, NC geometry provides a subtle way to spacetime quantization and the most promising feature is that it removes the short-distance divergences of the theory.\newline

  It may be noted that a study of the NC geometry inspired black hole is very important as it cures the usual problem encountered in the terminal phase of the classical black hole evaporation.  Further, the NC inspired black hole has no curvature singularity. It rather has de Sitter core at a short distance. Many authors have extended the classical black hole solution to find the exact NC inspired black holes \cite{Nicolini1}. In this paper, we study the effect of the NC geometry upon spherically symmetric accretion onto a NC geometry inspired Schwarzschild black hole. Interestingly, it turns out that the mass accretion rate $\dot{M}$ surprisingly decreases in comparison to the  Schwarzschild black hole \cite{Michel}.\newline
 
 The paper is organized as follow In Sec.~\ref{NC}, we review the NC inspired black hole and study its properties. In Sec.~\ref{eqn}, the analytic relativistic accretion onto a NC inspired Schwarzschild black hole model is appropriately developed. We calculate how the presence of a smeared mass object instead of point mass object would affect the mass accretion rate $\dot{M}$ of a gas onto a black hole. We also determine the analytic corrections to the critical radius, fluid critical velocity and the sound speed. We then obtain expressions for the asymptotic behavior of the fluid density and the temperature near the event horizon in Sec.~\ref{asym}. Finally the conclusion is found in Sec.\ref{conclusion}.

 We use the following values for the physical constants for numerical computations and plots:
 $c = 3.00 \times 10^{10} \mathsf{cm. s}^{-1}$,
 $G = 6.674 \times 10^{-8} \mathsf{cm}^{3}. \mathsf{g.}^{-1} \mathsf{s}^{-2}$,
 $k_B = 1.380 \times 10^{-16} \mathsf{erg. K}^{-1}$,
 $M = M_{\odot} = 1.989 \times 10^{33} \mathsf{g}$,
 $m_b = m_{p} =1.67 \times 10^{-24} \mathsf{g}$,
 $n_{\infty} = 1 \mathsf{cm}^{-3}$,
 $T_{\infty} = 10^4 \mathsf{K}$.\\
\section{Noncommutative geometry inspired black hole}\label{NC}
The analytic relativistic accretion
solution onto the  black hole by Michel \cite{Michel}
is generalized by considering a NC inspired Schwarzschild black hole. We begin with a brief review of a NC inspired Schwarzschild black hole and its properties (see \cite{Nicolini1} for further details). NC geometry is employed to implement the fuzziness of spacetime by replacing the position of Dirac's delta function by the spatially spreading distribution of $\rho$ \cite{Smailagic1, Smailagic2}. The choice of the mass density distribution is not in a general form. Rather, it explicitly depends upon the construction of the underlying NC geometry. Here we consider the widely studied spherically symmetric mass distribution~\cite{Nicolini1}, which is a Gaussian distribution with minimal width $\sqrt{\theta}$,
\beq\label{rho}
\rho_{\theta}(r)=\frac{M}{(4\pi\theta)^{3/2}}\exp(-r^2/{4\theta}).
\eeq
The particle mass $M$ is diffused throughout a region of linear size $\sqrt{\theta}$. Beyond the energy scale $1/\sqrt{\theta}$ spacetime becomes NC. For the Lorentz invariance and UV finiteness of NCQFT we must have the same NC parameter in all different planes. \\
For a Schwarzschild-like geometry $g_{tt}=-(g_{rr})^{-1}$ the energy momentum tensor constrained by $T^t_t=T^r_r=-\rho_{\theta}$ and the covariant conservation of $T^{\mu}_{\nu}$ ( $T^{\mu\nu}_{;\nu}=0$) lead to the solution
\beq\label{p}
T^{\theta}_{\theta}=-\rho_{\theta}-\frac{r}{2}\partial_{r}\rho_{\theta}.
\eeq
Therefore, under NC geometry a massive point source turns into self-gravitating, anisotropic fluid-type matter of density $\rho_{\theta}$ and pressure $P_r=-\rho_{\theta}$,  $P_{||} =-\rho_{\theta}-\frac{r}{2}\partial_{r}\rho_{\theta}$.
Furthermore, modification in the Newtonian potential in NC spacetime can be traced back to the change in the Green function (Feynman propagator) caused by the NCY of the spacetime coordinates \cite{A. Gruppuso}. The potential is related to the (tt) component of a fluctuation in the metric through $\phi=h_{tt}/2$, where
\begin{equation}
h_{tt}(x)=-8\pi G_N\int d^4x' G(x,x') T_{tt}(x'),
\end{equation} 
 $G(x,x')$ is the modified Feynman propagator in NC spacetime. After straightforward calculation we get the concrete expression of the potential,
\begin{equation}
\phi(r)=\frac{-GM}{r}erf\left(\frac{r}{2\sqrt{\theta}}\right),
\end{equation} 
which matches with the Newtonian potential at a large distance, and differs exponentially at a short distance. Here at $r=0$, we get
\beq
\phi(0)=-\frac{GM}{\sqrt{\pi\theta}}.\nn
\eeq
Hence, as expected, NC geometry regularizes the divergence at a short distance.
Solving the Einstein equation $G_{ab}=8\pi T_{ab}$ with the matter source of Eq.~(\ref{rho})  we find 
the line element 
\bea\label{line element1}
ds^2 &=& \Big(1-\frac{4M}{r\sqrt{\pi}}\gamma(3/2,r^2/{4\theta})\Big)dt^2\\ \nn
& &-\Big(1-\frac{4M}{r\sqrt{\pi}}\gamma(3/2,r^2/{4\theta})\Big)^{-1}dr^2\\ \nn
& &-r^2(d\vartheta^2+\sin^2(\vartheta)d\phi^2),
\eea
where $\gamma(3/2,r^2/{4\theta})$ is the lower incomplete gamma function,
\beq
\gamma(3/2,r^2/{4\theta})=\int_{0}^{r^2/4\theta}e^{-t}t^{1/2}dt. 
\eeq
The classical Schwarzschild solution can be recovered in the limit $r/\sqrt{\theta}\rightarrow \infty$. Again in the limit $r\rightarrow \infty$ we find the Minkowski solution. On the event horizon we have $g_{rr}(r_H)=0$, i.e. 
\begin{equation}\label{horizon}
r_H=\frac{4M}{\sqrt{\pi}} \gamma\left(3/2, r_H^2/4\theta\right)
\end{equation}
which can be written in terms of an upper incomplete gamma function
\beq\label{horizon1}
 r_H=2M\Big(1-\frac{2}{\sqrt{\pi}}\gamma^u(3/2,M^2/\theta)\Big).
\eeq
\begin{figure}
\centering
\includegraphics[scale=0.7]{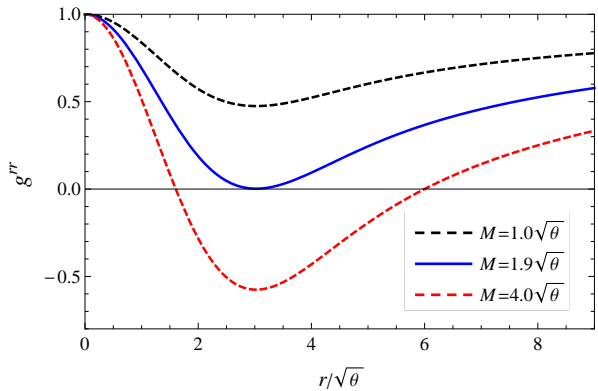} 
\caption{Plot of $-g_{tt}=g_{rr}^{-1}$ as a function of $r/\sqrt{\theta}$ for various values of $M/\sqrt{\theta}$. $M=1.0\sqrt{\theta}$ (\textit{black curve}) has no horizon, $M=1.9\sqrt{\theta}$ (\textit{blue curve}) has one degenerate horizon at $r_H=3.0\sqrt{\theta}$, and $M=4.0\sqrt{\theta}$ (\textit{red curve}) has two horizons}.
\label{fig:horizon}
\end{figure}
The numerical solution for the horizon is depicted in Fig.~\ref{fig:horizon}. It is clear that NCY modifies the behaviour of the horizon and it's radius. It turns out a critical mass $M_0$ can be found, such that a single degenerate horizon occurs for $M=M_0= 1.9\sqrt{\theta}$ with $r_H=3\sqrt{\theta}$. For $M>M_0$ there are two  horizons and whereas for $M<M_0$ there is no horizon. The larger root corresponds to the event horizon
of the  black hole and the smaller root is so called Cauchy (or inner) horizon. In fact spacetime curvature scalars are regular at $r=0$.
 In the large radius regime, we find
 \beq
 r_H=2M\Big(1-\frac{M}{\sqrt{\theta\pi}}e^{-M^2/\theta}\Big).
 \eeq
 The effect of NCY falls off exponentially and is of relevant significance, namely at small distance where spacetime fuzziness comes into picture. Hence, clearly the NC inspired black hole horizon radius $r_H$ is always less than or equal to the conventional black hole horizon radius $2M$, $r_H\leq 2M$.  
 
 The mass of the black hole can be obtained by integrating Eq.~(\ref{rho}), 
 \begin{equation}\label{mass}
m_{\theta}(r) = \int_{0}^{r} 4 \pi\; s^2\; \rho(s) \;ds = \frac{2M}{\sqrt{\pi}} \gamma(3/2,r^2/{4\theta}) = M f(r)
 \end{equation}
 and hence the line element can be conveniently put in the form
 \begin{eqnarray}\label{line element}
 ds^2 &=& \Big(1- \frac{2M f(r)}{r}\Big) dt^2 -\frac{1}{\Big(1- \frac{2M f(r)}{r}\Big)}dr^2 \\ \nn
 & &-r^2(d\vartheta^2+\sin^2(\vartheta)d\phi^2). 
 \end{eqnarray}
 where
 \beq
 f(r)=\frac{2}{\sqrt{\pi}}\gamma(3/2,r^2/{4\theta}),
 \eeq
 in the limit, $r/\sqrt{\theta}\rightarrow\infty$, $f(r)=1$.
 The metric of Eq.~(\ref{line element}) is useful to discuss the properties of accretion onto the NC inspired black hole, which is analyzed in the next section. 
\section{Accretion onto black hole}\label{eqn}
Here we generalize the results of the relativistic accretion onto the NC inspired Schwarzschild black hole and analyze how the NCY demands a transition to a supersonic flow in the solution. A black hole will capture the interstellar ambient matter present within a certain effective distance. Due to the spherical symmetry this accretion will be a steady state radial inflow of gas \cite{Michel, stbook}. In the low energy limit the quantum gravity effects (NCY of spacetime) can be neglected. Thus the gas is approximated as an isotropic, though inhomogeneous perfect fluid. The inhomogeneity in the fluid is due to the fact that near the accreting body the energy density of gas will be position dependent. The gas is described by the stress tensor
\begin{equation}
\label{energy-mom}
T^{\mu\nu} = \left( \rho + p \right)u^\mu u^\nu + p g^{\mu\nu},
\end{equation}
where  $\rho$ and $p$ are the fluid energy density and pressure in the comoving frame, respectively, and
\begin{equation}
\label{velocity}
u^\mu = \frac{d x^\mu }{d\tau},
\end{equation}
is the fluid's proper four-velocity defined with respect to the proper time $\tau$. In the comoving frame baryon number
density  $n$ and the  baryon number flux are related as $J^\mu  = n u^\mu$. The conservation of the baryon number flux
\begin{equation}
\label{baryoncons}
\nabla_\mu J^\mu = \nabla_\mu ( n u^\mu ) = 0,
\end{equation}
and the conservation of the stress tensor yield,
\begin{equation}
\label{momencons} \nabla_\mu T^{\mu}_{\nu} = 0.
\end{equation}
With the normalization condition $u_\mu u^\mu = -1$, and due to the restriction of spherically symmetric accretion the velocity components become $u^{\theta}=0, u^{\phi}=0$; $u^{\mu}=(u^0,u^1(=v(r)),0,0)$, so we have 
\begin{equation}
\label{zerocomp} u^0 = \left[\frac{v^2 + 1 - \frac{2Mf(r)}{r}}{\left( 1 - \frac{2Mf(r)}{r}
\right)^2}\right]^{1/2}.
\end{equation}
We rewrite Eq.~(\ref{baryoncons}) as
\begin{equation}
\label{baryoncons1}
\frac{1}{r^2} \frac{d}{dr} \left( r^2 n v \right)= 0.
\end{equation}
We now calculate the three-velocity of the accreting fluid for the distant observer, defined by $w=u^iu_i$,
\bea
w=\frac{u^1}{u^0}\left(\frac{1}{1-\frac{2Mf(r)}{r}}\right),\nn\\
\label{three velocity}
=\frac{v}{(1-\frac{2Mf(r)}{r}+v^2)^{1/2}}.
\eea
The assumption of spherical symmetry and a steady state radial inflow lead us to reformulate Eq.~(\ref{momencons}) in a simple form. The $\nu = 0$ component gives
\begin{equation}
\label{momencons0} \frac{1}{r^2} \frac{d}{dr} \left[ r^2 (\rho + p)
v \left(  1 - \frac{2Mf(r)}{r} +   v^2 \right)^{1/2} \right] = 0,
\end{equation}
and $\nu=1$ component can be simplified to
\begin{equation}
\label{momencons1}
v \frac{dv}{dr} = -  \left( \frac{ 1 - \frac{2Mf(r)}{r} + v^2  }{\rho + p} \right)\frac{dp}{dr} -\frac{Mf(r)}{r^2}+\frac{M}{r^2}\frac{df(r)}{dr}.
\end{equation}
The above equations are generalizations of the result obtained for the standard black hole.
\subsection{Accretion process}
The theoretical scrutiny of the accretion process in such a NC geometry can inject vitality into the research toward accretion onto a black hole.
We are considering the spherical steady state accretion onto a NC inspired Schwarzschild black hole of mass $M$. The boundary conditions at infinity are the same as the  Schwarzschild black hole \cite{Michel}. Since there is no entropy production for adiabatic fluid, the mass conservation equation is
\begin{equation}
\label{secondlaw}
T ds = 0 = d \left( \frac{\rho}{n}\right) + p~d\left(\frac{1}{n}\right),
\end{equation}
which may be rewritten
\begin{equation}
\label{slaw1}
\frac{d\rho}{dn} = \frac{\rho + p}{n}.
\end{equation}
A crucial physical entity in the accretion process is the speed of sound $a$ defined as \cite{stbook}
\begin{equation}
\label{soundspd}
a^2 \equiv \frac{dp}{d\rho} = \frac{dp}{dn} \frac{n}{\rho + p}.
\end{equation}
The baryon conservation given by Eq.~(\ref{baryoncons}) and the momentum conservation are
\beq
\label{rho4}
(\rho+p)u^{\mu}_{;\mu}+u^{\mu}\rho_{,\mu}=0,
\eeq
on inserting Eq.~(\ref{soundspd}) in Eqs.~(\ref{baryoncons}) and (\ref{rho4}) these are modifies to
\begin{align}
\label{bcon}
& \frac{v'}{v} + \frac{n'}{n}+\frac{2}{r} = 0, \\
\label{mcon1}
& \Big[vv' + a^{2}\left( 1 - \frac{2Mf(r)}{r}+ v^2\right) \frac{n'}{n}\nn \\ 
& +\frac{Mf(r)}{r^2}- \frac{M}{r}\frac{df(r)}{dr}\Big] = 0,
\end{align}
where a prime ($'$) denotes a spatial derivative, solving the coupled Eqs.~(\ref{bcon}) and (\ref{mcon1}) simultaneously:
\begin{subequations}\label{system}
\begin{align}
\label{system:1}
v' &= \frac{N_1}{N},  \\
\label{system:2}
n' &= -\frac{N_2}{N}.
\end{align}
\end{subequations}
These equation are known as Bondi equations \cite{Bondi}, with
\begin{subequations}\label{def}
\begin{align}
\label{def:1}
N_1 & = \frac{1}{nr^2} \left[ \left(1 - \frac{2Mf(r)}{r} + v^2 \right)2a^2r - Mf(r)+ Mrf'(r)  \right], \\
\label{def:2}
N_2 & = \frac{1}{vr^2} \left[2v^2r - Mf(r)+ Mrf'(r)  \right], \\
\label{def:3} N &= \frac{1}{vn} \left[ v^2 - \left( 1 - \frac{2Mf(r)}{r} + v^2
\right)a^2 \right].
\end{align}
\end{subequations}
For the same boundary conditions there are various solution classes but we are only interested in those for which the velocity increases monotonically from $v=0$ at $r=\infty$ to $v=1$ at $r=r_H$. For a large $r$ the flow is subsonic, i.e. $v < a$. Since the sound speed must be sub-luminal, i.e. $a^2 < 1$, this ensure $v^2 \ll
1$. Therefore, the denominator of Eq.~(\ref{system}) defined by Eq.~(\ref{def:3}) takes the form
\begin{equation}
N \approx \frac{v^2 - a^2}{vn}<0,
\label{N1}
\end{equation}
at the event horizon $r_H = 2Mf(r_H)$, again under the causality constraint $a^2<1$. Hence
\begin{equation}
N = \frac{v(1-a^2)}{n}>0.
\label{N2}
\end{equation}
Therefore, from Eqs.~(\ref{N1}) and (\ref{N2}), it is clear that the flow must pass through a critical point $r_H<r_s<\infty$, where $N=0$. The steady and continuous radial inflow demands $N_{1}=N_{2}=0$ at $r_s$, otherwise the turn-around point will occur in the trajectory and the solution will be double valued in either $r$ or $v$. This is nothing but the so-called {\it sonic
condition} and $r_s$ is called sonic radius. The solution of Eq.~(\ref{system}) governs the crossing of the critical point known as the transonic solution. At $r=r_s$ from the Eq. (\ref{def}) we find that
\begin{equation}
\label{sonic} v_{s}^{2} = \frac{a_{s}^{2}}{1 - a_{s}^{2}}\left[1-\frac{2Mf(r_{s})}{r_{s}}  \right]
= \frac{M}{2r_{s}}\left[f(r_{s})-r_{s}f'(r_{s}) \right],
\end{equation}
where $v_s \equiv v(r_s)$ and $a_s \equiv a(r_s)$.  The quantities
with a subscript $s$ are defined at the sonic
points of  the flow. It is clear from Eqs.~(\ref{three velocity}) and (\ref{sonic}) that at the critical point $r_s$ the accreting fluid speed will be $w_s=a_s$. For the standard Schwarzschild black hole  $f(r)=1$ we find the celebrated result of Michel's work:
\beq
\label{velocityC}
{v_s}^2=\frac{M}{2r_s}=\frac{a_s^2}{1+3a_s^2}.
\eeq
 It can be clearly seen that the critical velocity
in this model is modified by the NC geometry parameter. The physically acceptable solution $v_{s}^{2}> 0 $ is guaranteed from Eq.~(\ref{sonic}).

The Bondi mass accretion rate can be obtained by integrating Eq.~(\ref{baryoncons1}) over a 4-dimensional volume and multiplying this
by $m_b$, the mass of each baryon, to get
\begin{equation}
\label{accrate} \dot{M} = 4 \pi r^2 m_{b}nv,
\end{equation}
where $\dot{M}$ is the mass accretion rate having the dimension of mass per unit time. The most influencing equations for studying accretion are Eqs.~(\ref{baryoncons1}) and (\ref{momencons0}). Collectively these yield
\begin{equation}
\label{bernoulli} \left( \frac{\rho + p}{n} \right)^2 \left( 1 -
\frac{2Mf(r)}{r} + v^2 \right) = \left( \frac{\rho_{\infty} +
p_{\infty}}{n_{\infty}}\right)^2,
\end{equation}
which is the improved relativistic Bernoulli equation for the
steady state accretion onto NC inspired black hole where the back-reaction of fluid is ignored. The mathematical expressions Eqs.~(\ref{accrate}) and (\ref{bernoulli}) are the characteristic equations of accretion with parameter $\theta$. In the
limit  $r/\sqrt{\theta}\rightarrow \infty$, our results reduce to the standard Schwarzschild black hole obtained in \cite{Michel,stbook}.\newline
The constraint of stationary accretion is self consistent with the following: (1) the accretion fluid is light-weight, (2) the rate of the black hole mass growth is slow. The back-reaction of the accreting matter onto spacetime keeps most of the features of a sonic point intact. The dynamics of a system with back-reaction can easily be understood through the investigation of the steady flow with the back-reaction being ignored \cite{malec, Karkowski}. If the test fluid makes only a slight contribution to the total asymptotic mass of system (black hole + fluid) we can safely ignore the back-reaction. Mathematically it gives
\begin{subequations}\label{back reaction}
\begin{align}
\label{back reaction:1}
& 4\pi\int_{R>2M} \rho r^2 dr<<m_*  \\
\label{back reaction:2}
& \rho<<\frac{m_*}{\pi r_s^3} 
\end{align}
\end{subequations}
where $m_*$ is the asymptotic mass of the system $(m_*=M_{BH}+m_{fluid})$. Therefore, back-reaction is of importance only when the accreting fluid is heavier than the central black hole defying Eq.~(\ref{back reaction}) where the metric depends upon the infalling matter and the back-reaction should be taken into account. Therefore, dealing with the ordinary matter present in an interstellar medium we can neglect the effect of back-reaction during its accretion onto a sufficiently massive black hole. On the other hand, the NC geometry of the underlying spacetime only modifies the source term in Einstein field equation while keeping the Einstein tensor intact. The reasoning behind this is that NCY is an intrinsic property of spacetime rather than an imposed condition.While studying accretion we have already taken account the modified spacetime metric due to NCY. Therefore, the fuzzy structure of spacetime at the Planck level will not further alter the back-reaction condition for the accretion process.
\subsection{The polytropic solution}

We consider the case when the critical point lies outside the outer horizon of the improved Schwarzschild black hole. To calculate $\dot{M}$ following Bondi \cite{Bondi} and Michel \cite{Michel} we
introduce a polytropic equation of state for a fluid
\begin{equation}
\label{eos} p = K n^{\Gamma},
\end{equation}
where $K$ and the adiabatic index $\Gamma$ are constants. Using Eq. (\ref{eos}) in (\ref{secondlaw}), integration leads to 
\begin{equation}
\label{rho1} \rho = \frac{K}{\Gamma -1}n^{\Gamma} + m_{b}n,
\end{equation} 
where $m_b$ is an integration constant which is evaluated comparing with $\rho=m_b n+\epsilon$ with $\epsilon$ as the internal density.
Hence Eqs.~(\ref{eos}) and (\ref{rho1}) with Eq.~(\ref{soundspd}) give
\begin{equation}
 \label{inter}
 \Gamma Kn^{\Gamma-1}=\frac{a^2m_{b}}{(1-\frac{a^2}{\Gamma-1})}.
\end{equation}
Further, substituting Eqs.~(\ref{rho1}) and (\ref{inter}) in the Bernoulli equation (\ref{bernoulli}), we obtain
\begin{align}
\label{ber1}
&\left(1 + \frac{a^{2}}{\Gamma - 1 - a^{2} }\right)^2 \left( 1  - \frac{2Mf(r)}{r}+ v^2 \right) \nonumber \\
& = \left(1 +  \frac{a_{\infty}^{2}}{\Gamma - 1 -
a_{\infty}^{2}}\right)^2.
\end{align}
This equation will also be true at a critical radius $r_s$. Using Eq.~(\ref{sonic}) we find
\begin{equation}
\label{berson}  \left( 1 -\frac{a_{s}^{2}}{\Gamma -1} \right)^2\left(1 -\frac{2Mf(r_s)}{r_s} \right)^{-1}(1-a_s^2) = \left( 1 -
\frac{a_{\infty}^{2}}{\Gamma - 1} \right)^2.
\end{equation}

For large but finite values of  $r$, i.e. $r \geq r_s$, it is likely that baryons will
be non-relativistic. In this regime we should have $a \leq a_s \ll 1$.
Expanding Eq.~(\ref{berson}) up to the second order in $a_s$ and
$a_{\infty}$ we find
\begin{align}
\label{sonsound}
\Big(\frac{\Gamma+1}{\Gamma-1}\Big)a_{s}^{2}= \frac{2a_{\infty}^{2}}{\Gamma - 1}+\frac{2Mf(r_s)}{r_s}\left[\frac{(\Gamma-1)-2a_{\infty}^{2}}{\Gamma-1}\right].
\end{align}
Clearly, for $a_{s}^2>0$, we have 
\beq
r_s>2Mf(r_s)\Rightarrow r_s>r_H\left[\frac{f(r_s)}{f(r_H)}\right].
\eeq
In the limit $r/\sqrt{\theta}\rightarrow \infty$, from Eqs.~(\ref{velocityC}) and (\ref{ber1}), we obtain
\beq
\label{sonicC}
a_{s}^2=\frac{2}{5-3\Gamma}a_{\infty}^{2}.
\eeq
We note from Eqs.~(\ref{sonsound}) and (\ref{sonicC}) that the sound speed at the critical point is enhanced due to the NC effect. 
From  Eqs.~(\ref{sonic}) and (\ref{sonsound}) we have
\beq
\label{sonrad}
r_s=\frac{(5-3\Gamma)Mf(r_s)\left[2a_{\infty}^2-(\Gamma-1)\right]}{4a_{\infty}^2(1-\Gamma)+Mf'(r_s)(\Gamma+1)\left[2a_{\infty}^2-(\Gamma-1)\right]},
\eeq
this is the critical radius for a polytropic fluid of an adiabatic index $\Gamma$.
For $a_{\infty}^2/(\Gamma-1) \ll 1$ we get 
\beq
r_s\approx\frac{(5-3\Gamma)f(r_s)}{(\Gamma+1)f'(r_s)}
\eeq
Under commutative consideration $r/\sqrt{\theta}\rightarrow\infty$ we get the standard result for the Schwarzschild black hole accretion, namely 
\beq
\label{radiusC}
r_s=\frac{(5-3\Gamma)}{4}\frac{M}{a_{\infty}^2}.
\eeq 
It may be noted from Eqs.~(\ref{sonrad}) and (\ref{radiusC}) that the critical radius decreases in the case of NC inspired black hole. This will put an influential impact upon the mass accretion rate and other characteristics of the  accretion. 
Under the limiting case $a^{2}/(\Gamma-1) \ll 1$, Eq.~(\ref{inter}) will take the form
\begin{equation}
\label{barden1} \frac{n}{n_{\infty}} \approx \left(
\frac{a}{a_{\infty}} \right)^{2/(\Gamma-1)}.
\end{equation}

The accretion rate $\dot{M}$ described by (\ref{accrate}) must also hold for $r=r_s$. At a critical
point the Bondi accretion rate turns out to be
\beq
\label{massRate}
\dot{M} = 4\pi r_{s}^2 m_{b}n_{s}v_{s}.
\eeq
By virtue of Eqs. (\ref{sonic}), (\ref{sonsound}), (\ref{sonrad})
and (\ref{barden1}) the accretion rate becomes
\begin{equation}
\label{accrate1} \dot{M} = 2\pi\left(\frac{a_s}{a_{\infty}}\right)^{2/(\Gamma-1)}
r_s\left(f[r_s]-r_s f'[r_s]\right)(m_b n_{\infty} M).
\end{equation}

\begin{figure}
\includegraphics[scale=0.85]{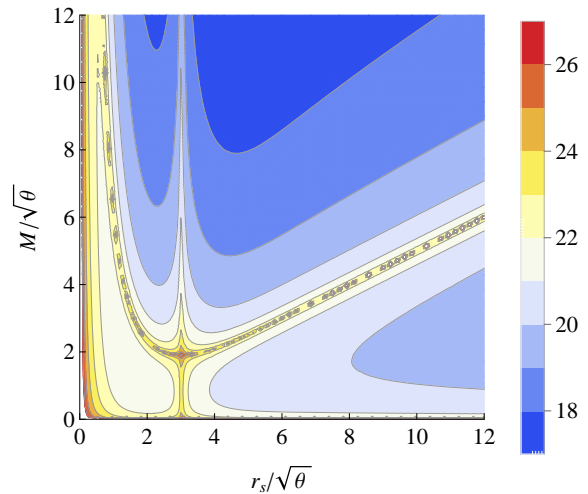}
\caption{ Mass accretion rate for accreting fluid of $\Gamma=1.5$ is plotted along the \textit{color axes} as a function of $r_s/\sqrt{\theta}$ and $M/\sqrt{\theta}$.  }
\label{Fig:massacrre}
\end{figure}
The accretion rate has a mass dependency as $\dot{M} \sim
M^2$ which is similar to that of the  Newtonian model \cite{Bondi} as well as
  the relativistic case \cite{Michel,stbook}. If no effects of NC geometry are taken into account substituting Eqs.~(\ref{velocityC}), (\ref{sonicC}) and (\ref{radiusC}) into Eq.~(\ref{massRate}), this gives the well-known result derived in  \cite{Michel,stbook},
\beq
\dot{M} = 4\pi\lambda_s(m_b n_{\infty} M^2 a_{\infty}^{-3}),
\eeq
with 
\beq
\lambda_s=\left(\frac{1}{2}\right)^{(\Gamma+1)/2(\Gamma-1)}\left(\frac{5-3\Gamma}{4}\right)^{-(5-3\Gamma)/2(\Gamma-1)}.
\eeq
  In Fig.~\ref{Fig:massacrre} we plotted the logarithm of
the accretion rate $\dot{M}$ against the $r_s/\sqrt{\theta}$ and $M/\sqrt{\theta}$.
\subsection{Some numerical results}
The most essential equations characterizing the accretion of fluid are Eqs.~(\ref{baryoncons1}) and (\ref{ber1}). Even though they look simple, these equations are difficult to solve analytically. To get a numerical solution \cite{charge1} it is a feasible idea to work with dimensionless variables, i.e. the radial distance ($x=r/2M$) and the particle number density ($y=n/n_{\infty}$). In terms of these newly defined variables Eq.\ (\ref{ber1}) can be re-expressed as
\begin{align}
\label{num1}
&\left(1 + \frac{a_{\infty}^2}{\Gamma - 1 }y^{\Gamma-1}\right)^2 \left( 1  - \frac{f(2Mx)}{x} + v^2 \right)  \nonumber \\
& = \left(1 +  \frac{a_{\infty}^{2}}{\Gamma - 1}\right)^2.
\end{align}
 \begin{figure*}
 \begin{tabular}{c c c}
 \includegraphics[scale=0.65]{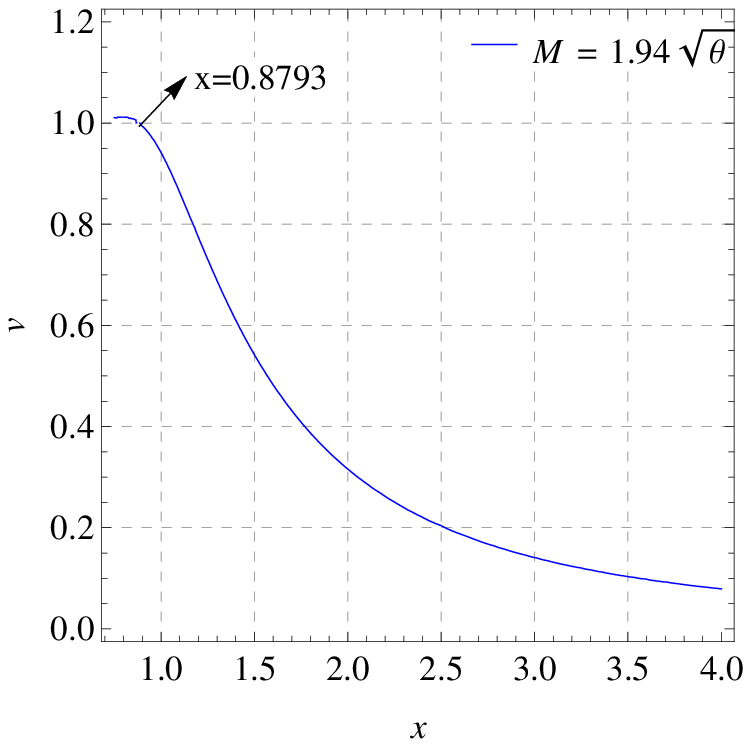}&
  \includegraphics[scale=0.65]{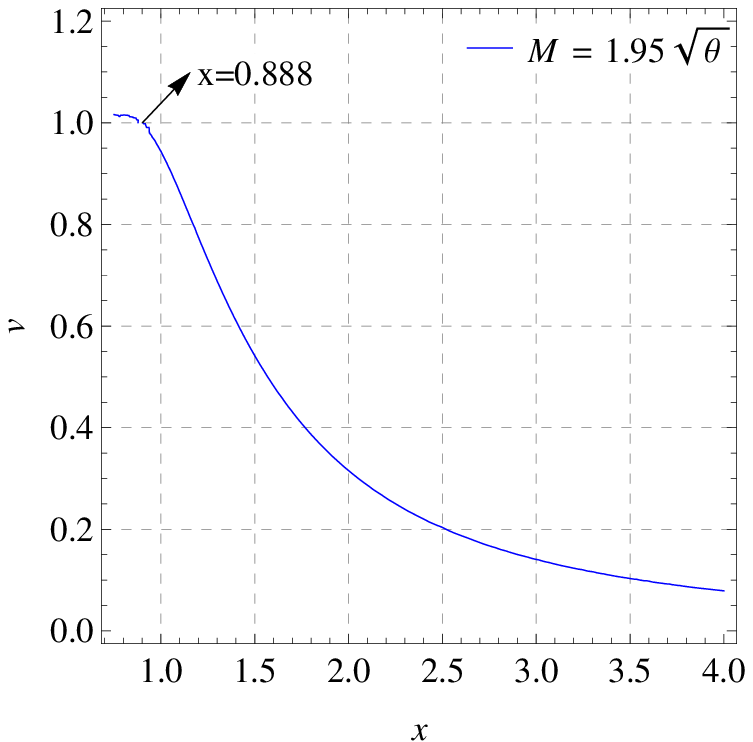}&
 \includegraphics[scale=0.65]{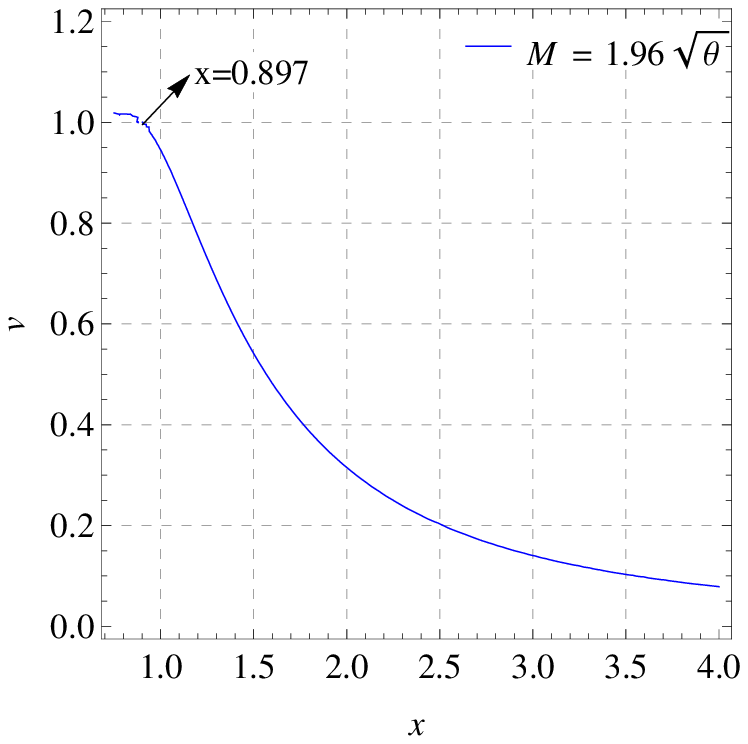}\\
 
 \includegraphics[scale=0.65]{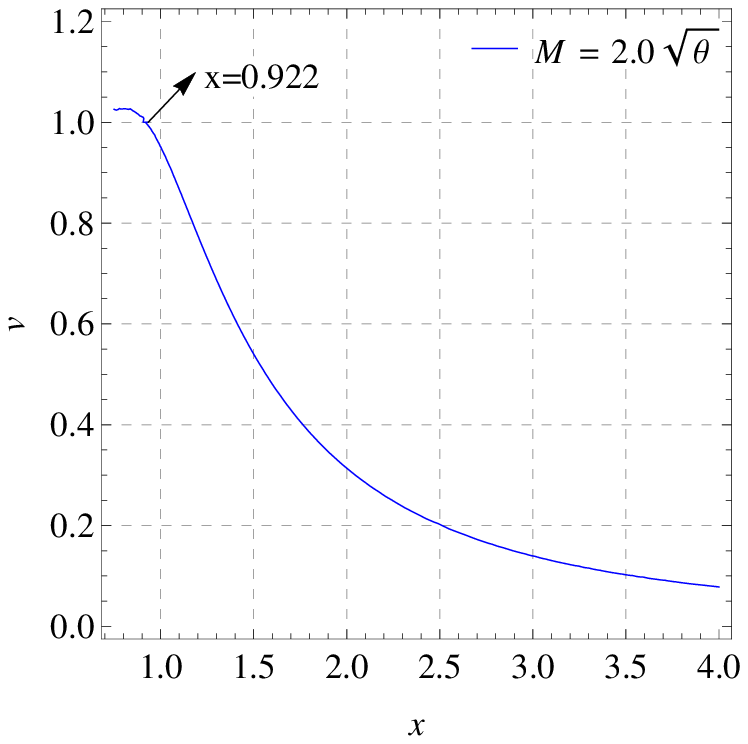}&
  \includegraphics[scale=0.65]{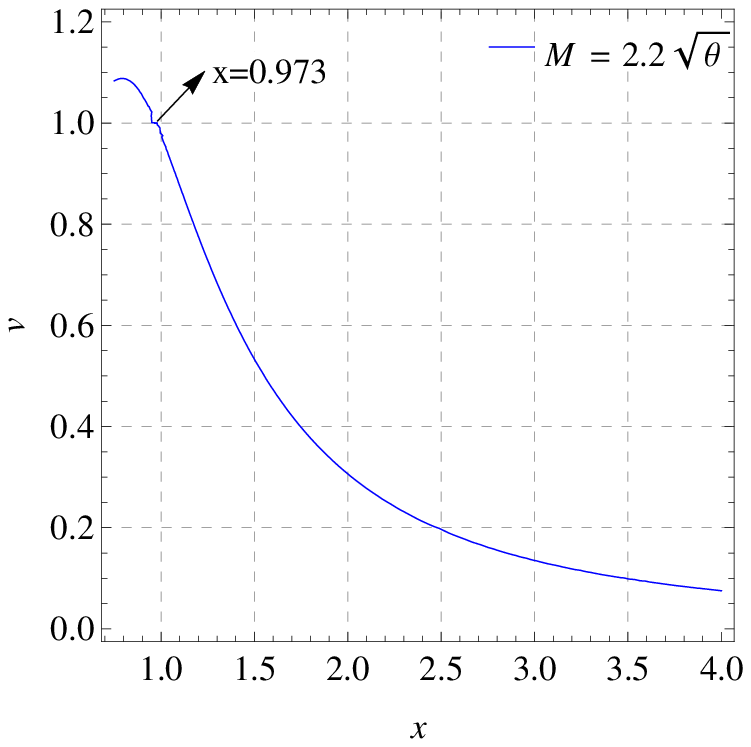}&
 \includegraphics[scale=0.65]{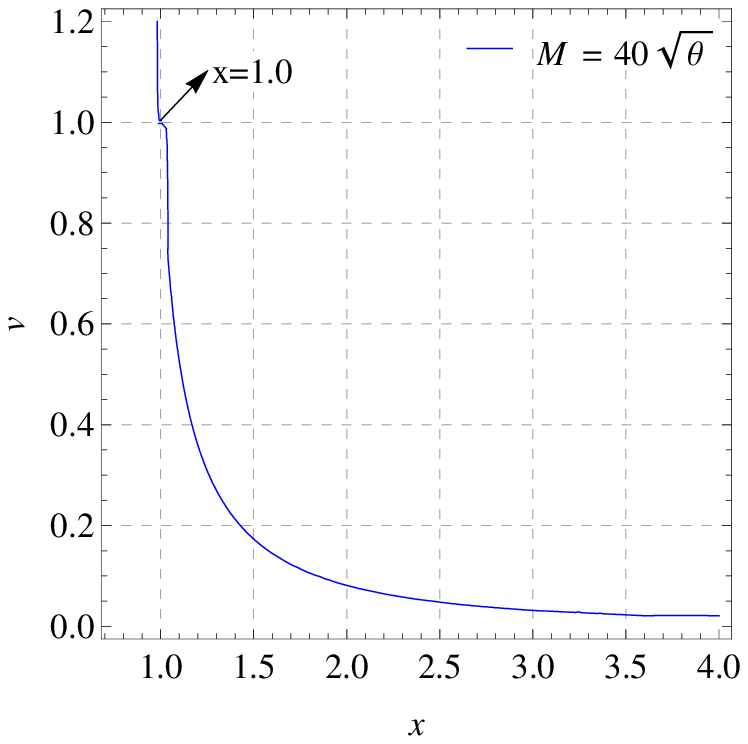}
  \end{tabular}
  \caption{Plot showing the radial velocity profile $v$ for a fluid $\Gamma=1.5 $ accreting onto the noncommutative inspired \textit{black hole} as a function of the dimensionless radius $x$.}
\label{Fig:velocity}
 \end{figure*}
While the baryon conservation equation (\ref{baryoncons1}) will take the form 
\beq
nvr^2=n_s v_s r_s^2, \nn 
\eeq
it can be re-cast as
\beq
yv=\left(\frac{x_s}{x}\right)^2\left(\frac{n_s}{n_\infty}\right) v_s.
\eeq
Following Eq.~(\ref{barden1}), we get
\beq
\label{num2}
yv=\left(\frac{x_s}{x}\right)^2\left(\frac{a_s}{a_\infty}\right)^{2/(\Gamma-1)} v_s.
\eeq
\\
Now we are left with the non-linear equations~(\ref{num1}) and (\ref{num2}), which can easily be solved numerically for both the fluid velocity $ v$ and the compression ratio $y$. The velocity  profile for the radial inflow of the accreting fluid $\Gamma=1.5$ as a function
 of the dimensionless variable $x$ is plotted in Fig. \ref {Fig:velocity}.

 \begin{figure*}
 \begin{tabular}{c c}
 \includegraphics[scale=0.8]{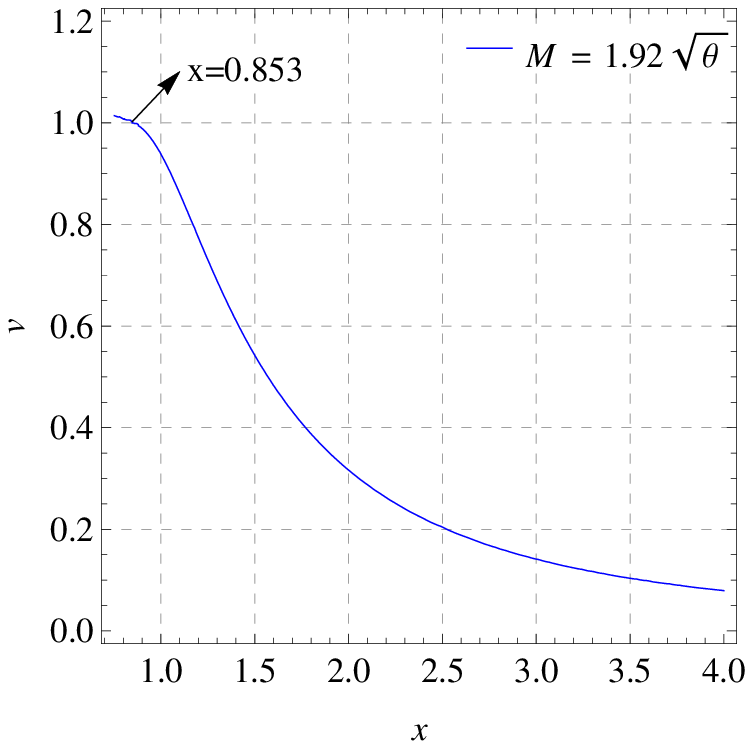}&
 \includegraphics[scale=0.8]{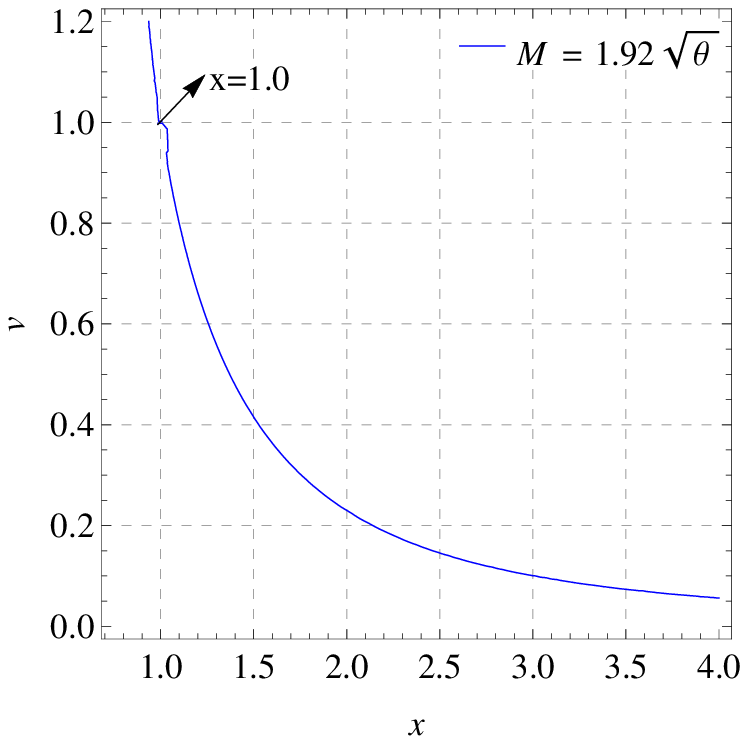}
  \end{tabular}
  \caption{The radial velocity profile $v$ for a fluid $\Gamma=1.5 $ accreting onto the \textit{black hole} as a function of a dimensionless radius $x$. (\textit{Left}) NC inspired black hole. (\textit{Right}) Conventional black hole.}
\label{Fig:velocity1}
 \end{figure*}
The event horizon $r_H$ for the NC inspired black hole has a profound dependency upon $\theta$. From Fig.~\ref{Fig:velocity} it is evident that accreting fluid crosses the event horizon with the speed of light. For example, a black hole of mass $M=1.92\sqrt{\theta} $ has an outer horizon radius $r_H=3.276\sqrt{\theta}< 2M $, a mass $M=1.94\sqrt{\theta} $ has $r_H=3.4115\sqrt{\theta} $, a mass $M=1.96\sqrt{\theta}$ has $r_H=3.5154\sqrt{\theta}$, a mass $M=2.0\sqrt{\theta}$ has $r_H=3.68\sqrt{\theta}$. For $M=40\sqrt{\theta}$ the horizon radius is $r_H\approx80\sqrt{\theta}(=2M)$. Clearly, NC effects are subjugate for low mass regime. In Fig.~\ref{Fig:velocity1} we compare the radial velocity profile of accreting gas for NC inspired black hole with the conventional black hole.

%
%
 
\begin{figure}
\centering
\includegraphics[scale=0.9]{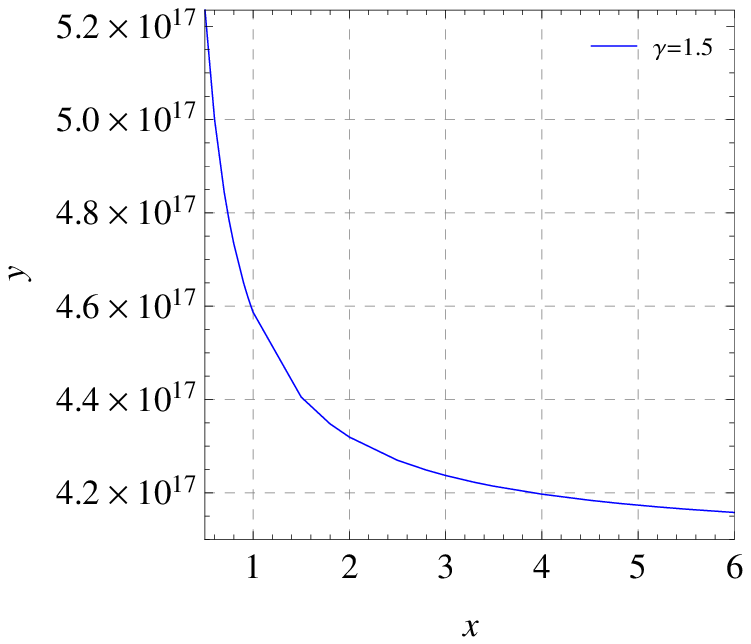} 
\caption{The compression factor ($y$) for a fluid $\Gamma=1.5$ accreting onto the \textit{black hole} with $M=40\sqrt{\theta}$ as a function of the dimensionless radius $x$.}
\label{Fig:compression}
\end{figure}

The variation of the compression ratio $y$ as a function of the radial coordinate for an accreting gas with $\Gamma = 1.5$ is shown in Fig.~\ref{Fig:compression}.
We have shown that the black hole mass is a dynamical quantity governed by the modified Bondi accretion rate. The Total integrated flux due to the surface luminosity $L_{\nu}$ reads
\beq
F_{\nu}=\frac{L_{\nu}}{4\pi d_L^2},
\eeq
where the surface luminosity is proportional to the accretion rate $L_{\nu} \propto \dot{M}$. This integrated flux is modified due to the NC behavior of the black hole. Therefore an extra integrated flux as compared to conventional black hole is
\beq
F=\frac{F_{\nu}-F_{\nu 0}}{F_{\nu 0}},
\eeq 
where $F_{\nu}$ and $F_{\nu 0}$ are the flux for the NC inspired black hole and for the conventional black hole, respectively. For $M>>M_0$, we have $F_{\nu 0}>> F_{\nu }$. Therefore $F\sim -1$. This is due to the fact that the accretion radius decreases due to NC effects and so does the accretion rate.

\section{Asymptotic behavior}\label{asym}
In the previous section we mainly discussed the characteristic of the accretion for the case where the critical point is far away from the outer horizon, i.e. $r_s \gg r_H$. In this section we will explore the other possibilities, $r_H < r
\ll r_s$ at the event horizon $r = r_H$.\newline
 The radial inflow solution is transonic in nature. Therefore for the case $r<r_s$, the fluid will be supersonic $v>a$. Near the horizon we can safely assume that
\begin{equation}
\label{supsonic}
v^2 \approx \frac{2Mf(r)}{r}, \; \qquad \Gamma \neq \frac{5}{3}. 
\end{equation}
Similarly the gas compression ratio will also get altered by using Eqs.~(\ref{baryoncons1}) and (\ref{barden1}) as
\bea
\label{gascomp}
\frac{n(r)}{n_{\infty}}&\approx &\left(\frac{n_s}{n_\infty}\right)\frac{r_s^2 v_s}{\sqrt{2Mf(r)r^3}},\\
\frac{n_s}{n_\infty}&=&\Big[\frac{1}{(\Gamma+1)}{2r_s-\frac{2Mf(r_s)}{a_\infty^2}}(2a_{\infty}^2+(\Gamma-1))\Big]^{1/(\Gamma-1)}.
\eea
For large $r$ values, $f(r)\approx 1$. Therefore the accretion situation will not much differ from the standard commutative accretion. However, at the small distance $f(r)\ll 1$ and hence number density at distance $r$ will get modified compared to the standard spherical accretion. 
Assuming that a gas follows the fundamental thermodynamic $p = nk_{B}T$, we find the adiabatic temperature profile, using Eqs.~(\ref{eos}) and (\ref{gascomp}):
\begin{equation}
\label{temp}
\frac{T(r)}{T_{\infty}} = \left[\frac{n(r)}{n_{\infty}}\right]^{\Gamma-1}.
\end{equation}
Next, at the event horizon of NC inspired black hole we have $r_H=2Mf(r_H)$. At a distance less than the Bondi radius the gas is always supersonic. Near the horizon the gas is always relativistic and rationally we can take $v_{H}^{2} \equiv v^2(r_H) \approx 1 $. Consequently, the gas compression ratio gets modified
\begin{equation}
\label{horcomp}
\frac{n_H}{n_{\infty}} \approx \left(\frac{n_s}{n_\infty}\right)\frac{r_s^2 v_s}{{r_H}^2}.
\end{equation}
The effect of NC geometry at the black hole horizon can be noticed from the fact that for a low mass black hole the horizon radius decreases due to the NC effects, which will significantly change the number density at the horizon. 
 
%
%
%

\section{Conclusions}\label{conclusion}
In recent years a flurry of activities toward research concerning NC inspired black hole physics could be witnessed. It can be asserted that, even though the NC structure of the spacetime is one of the exotic outcomes of the string theory, the paradigm of the NC geometry is more general. The NC geometry provides an effective framework to study short-distance spacetime dynamics. It turns out that a NC inspired Schwarzschild black hole smoothly interpolates between a de Sitter core around the origin and ordinary Schwarzschild black hole at a large distance. It has exquisite effects on black holes. Namely it turns an ordinary Schwarzschild black hole into a regular black hole with two horizons. Furthermore there is a compelling modification in the late stage evaporation of a black hole. Furthermore, accretion of the matter onto the black hole is one of the viable idea to explain the ultra high energy output from the active galactic nuclei and the quasars.   The parameter $\theta$ occurring in NC geometry suggests the existence of a very small scale leading to black holes that are actually microscopic. It is possible that quantum primordial black holes were created in the early Universe (or big bang), or possibly through subsequent phase transitions. They might be observed by astrophysicists in the near future through the particles they are expected to emit by Hawking radiation. The higher dimensions predict that micro black holes could be formed at energies as low as the TeV range, which are available in particle accelerators.  

This paper deals with a basic model of a steady state spherical accretion of a polytropic fluid onto a NC inspired Schwarzschild black hole. Owing to the low accretion rate for such a black hole we can safely ignore the back-reaction effect of matter onto the black hole. The resultant radial inflow of fluid is transonic in character even for conventional black hole \cite{Bondi}. But other accretion parameters such as the critical radius $r_s$, the sonic speed $a_s$, the accretion rate $\dot{M}$, and other thermodynamic quantities such as the gas compression ratio $y$ and the temperature at various distances from the horizon $T(r)$ get modified substantially due to NCY. Even a critical point located away from the outer horizon of a black hole has a subtle dependency on the NC parameter $\theta$. Due to NC effects the critical radius seems to decrease when compared compare to the conventional black hole. On the other hand, the sound speed at the critical point increases. Thus, despite the NCY complexity, we have determined analytically the critical radius, the critical fluid velocity and the sound speed and subsequently the mass accretion rate. We then obtained expressions for the asymptotic behavior of the fluid density and temperature near the event horizon.
 Hence, in this sense, we may conclude that the steady state spherical solution of accretion onto the Schwarzschild black hole is
stable.  In the limit $r/\sqrt{\theta} \rightarrow \infty$ the results obtained here are reduced exactly to vis-$\grave{a}$-vis 'a vis those obtained by \cite{Michel}.

\section*{Acknowledgements}

S.G.G. would like to thank SERB-DST Research Project Grant No. SB/S2/HEP-008/2014 and the DST INDO-SA bilateral project DST/INT/South Africa/P-06/2016 and also IUCAA, Pune, for the hospitality while this work was being done. R.K. would like to thank UGC for providing the Junior Research Fellowship.

\textbf{Note added in proof:} After this work was completed, we learned of a similar work by Biplab et al. \cite{Paik:2017wcy}, which appeared on 30 March 2017 in the arXiv .

\end{document}